\documentclass[12pt]{article} 
\usepackage{graphicx}

\begin{document}

  \vspace*{1.1cm} 
  \begin{center} 
  {\LARGE \bf The vacuum state of quantum gravity contains large virtual masses} 
  \end{center} 

  \begin{center} 
  \vskip 10pt 
  Giovanni Modanese\footnote{e-mail address: 
  giovanni.modanese@unibz.it}
  \vskip 5pt
    {\it University of Bolzano -- Logistics and Production Engineering \\
  Via Sernesi 1, 39100 Bolzano, Italy}
  \end{center} 

  \baselineskip=.215in 
    
\begin{abstract}
In the functional integral approach to quantum gravity, the quantum configurations are usually treated to order $\hbar$ through a stationary phase approximation around the saddle point of the action where spacetime is flat. We show that from this point a ``level line" in functional space departs, which comprises a family of static non-flat metrics with zero scalar curvature, depending on a continuous mass parameter. Furthermore, each of these metrics can be perturbed by an arbitrary function in such a way to still satisfy the condition $\int \sqrt{g}Rd^4x=0$. We thus find a set of zero-modes of the gravitational action which has non-vanishing measure in the functional space. These metrics will contribute to the functional integral as vacuum fluctuations, on the same footing as those near the saddle point.

\medskip

\noindent 04.20.-q Classical general relativity.

\noindent 04.60.-m Quantum gravity.

\medskip


\end{abstract}

\section{Introduction}

The functional-integral approach to quantum gravity was initiated in the 1970s and early 1980s by the monumental works of Hawking, Gibbons, Coleman and others \cite{eqg}. Quantum effects at first order in $\hbar$ were computed in the stationary phase approximation through the zeta-function regularization technique. Hawking also tackled the problem of the un-definiteness of the gravitational action and showed that the action is unbounded from below, due to conformal modes which can be integrated away. The non-conformal part of the action was assumed to be positive semi-definite and this conjecture was then proved in several special cases \cite{positive}. Later, the un-definiteness of the action and the continuation to imaginary time were studied by Greensite, Mazur and Mottola and others \cite{green}. Wetterich \cite{wet} found that the effective action is always un-defined in sign. 

In the 1990s, Hamber and Williams implemented numerically a discrete version of Euclidean quantum gravity on a Regge lattice and computed non-perturbative quantities like the effective coupling \cite{hw}. The Regge lattice approach also gives an insight into the cosmological constant problem, showing that a space with average null curvature can emerge dynamically from a full quantum theory which includes strong (gravitational-only) vacuum fluctuations. 

In this work we shall consider pure gravity without any sources or matter fields, the action being given by the Einstein-Hilbert term plus the York-Hawking-Gibbons boundary term. We are thus looking at the vacuum state of quantum gravity. Our main result concerns the existence of a class of static non-flat metrics with zero action, which should be taken into account when computing physical quantities in the functional-integral approach. 

In our previous work \cite{prev} we found explicit expressions for these ``zero-modes" in the weak field approximation. They owe their existence to the non-positivity of the Einstein-Hilbert lagrangian $\sqrt{g}R$. This feature makes possible a cancellation between positive and negative contributions to the integral of $\sqrt{g}R$, arising from distinct spacetime regions.

\begin{figure}
  \begin{center}
    \includegraphics{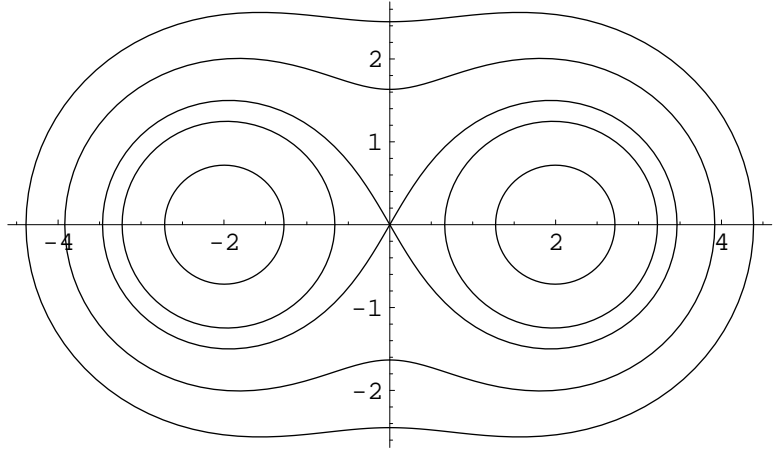}
  \end{center}
  \caption{Example of action level lines near a saddle point in two dimensions. A level line exists, where the action has the same value it takes at the saddle point.}
  \label{saddle}
\end{figure}

Here we obtain a generalization of the zero-modes to the strong field case. The situation can be understood through a finite-dimensional analogy. Imagine a system with an action which is not positive-definite. The stationary point corresponding to the classical theory will be a saddle point. The quantum configurations to order $\hbar$ span a small area around the saddle point. We can possibly find a ``level line" (Fig.\ \ref{saddle}) where the lagrangian has the same value it takes at the saddle point. Starting from the classical configuration at the saddle point, the system can evolve along this level line. Such lines do not exist, of course, for systems with positive-definite action. 

One should correctly object that level lines are subsets of null measure and do not contribute to an integration over the full space. But here the analogy fails to account for the complexity of a functional space. We shall prove that our strong-field zero-modes constitute a full-dimensional subset of the functional space of smooth, non-singular metrics, because they can be perturbed on an infinite-dimensional functional basis of the form $\{\sin(2\pi nx),\cos(2\pi mx)\}$. The oscillating weight factor $\exp(iS/\hbar)$ in the functional integral (where $S$ includes the boundary term) is constant over this set.

The next question is: how can we take these configurations into account for the computation of physical averages? This is a very difficult task, and a problem in functional analysis which goes well beyond the traditional stationary-phase approximation. From the physical point of view, however, it seems clear that such strong vacuum fluctuations should make feel their presence in some way. One can envisage, for instance, some kind of gravitational Casimir effect, where the variation of external constraints on the fluctuations results in observable vacuum forces.

In Sect.\ \ref{search} we use a ``virtual source" method to search for zero-modes of the action. In the context of wormholes physics \cite{visser,wh} this method is also called ``reverse solution of the Einstein equations"; it consists in finding a source which generates a metric with some desired features. In our case, the desired metric is not easy to define, because it must obey the non-linear integral condition $\int \sqrt{g}R d^4x=0$. In the weak-field approximation this translates, up to terms of order $G^2$, to the condition $\int T_{\mu \nu} \delta^{\mu \nu} d^4x=0$. 

While in wormholes physics it is necessary to minimize and justify in some way the recourse to un-physical matter, here we are free to introduce any source, provided it satisfies the Bianchi identities. For weak fields it is sufficient to choose $T_{ii}=0$, leaving only the $T_{00}$ component, and build a virtual source satisfying the ``dipolar" condition $\int T_{00} d^4x=0$. This can be made, for instance, by a positive-density core surrounded by a negative shell \cite{prev}. 

In the strong field case, however, the pressure components of $T_{\mu \nu}$ are essential in order to satisfy the Bianchi identity; from numerical solutions of the Oppenheimer-Volkoff equations, it turns out that the pressure tends to diverge in the negative-density shell, even beyond the point where the mass-energy density has been cut off. This way of looking for quantum configurations, by introducing unusual sources in the classical equations, seems therefore to be ineffective. 

In Sect.\ \ref{direct} we take a different, more straightforward approach. We consider the most general static symmetric metric and try to impose directly the zero-action condition.
We obtain a closed solution in the special case of constant $g_{00}$. In this case, the lagrangian density $\sqrt{g}R$ has a relatively simple expression in terms of $g_{rr}$, and we can invert this as a first-order differential equation for $g_{rr}$ in terms of $\sqrt{g}R$. Setting $R=0$, we obtain a family of non-flat metrics with zero scalar curvature, depending on a continuous parameter (the ``level line"). Replacing $\sqrt{g}R$ with an arbitrary periodic function with zero average, we obtain the more general zero-modes mentioned above.

\section{Search for zero-action modes with the virtual source method}
\label{search}

\subsection{Weak field}
\label{weak}

To first order in $G$, the
field $h_{\mu \nu}(x)$ generated by a mass-energy
distribution $T_{\mu \nu}(x)$ is given 
by an integral of the field propagator 
$P_{\mu \nu \rho \sigma}(x,y)$ over the source:
	\begin{equation}
	h_{\mu \nu}(x) = \int d^4y 
	P_{\mu \nu \rho \sigma}(x,y) T^{\rho \sigma}(y)
\label{campo}
\end{equation}

In Feynman gauge the propagator 
is given, with our conventions on the 
metric signature, by
	\begin{equation}
	P_{\mu \nu \rho \sigma}(x,y) = \frac{2G}{\pi}   
	\frac{\eta_{\mu \rho}\eta_{\nu \sigma} +
	\eta_{\mu \sigma}\eta_{\nu \rho} -
	\eta_{\mu \nu}\eta_{\rho \sigma}}
	{(x-y)^2 + i\varepsilon}
\end{equation}

Consider a static source in which only the component $T^{00}$ is non-zero.
To order $G$, such a source automatically satisfies the Bianchi identities.
The field is
	\begin{eqnarray}
	h_{\mu \nu}({\bf x}) & = & 
	\int_{-\infty}^{+\infty} dy_0 \int d^3y 
	T^{00}({\bf y}) P_{\mu \nu 00}(x,y) \nonumber \\
	& = & \frac{2G}{\pi} 
	(2\eta_{\mu 0} \eta_{\nu 0} - 
	\eta_{\mu \nu} \eta_{00})
	\int_{-\infty}^{+\infty} dy_0 \int d^3y 
	\frac{T^{00}({\bf y})}{(x_0-y_0)^2 -
	({\bf x}-{\bf y})^2 + i\varepsilon} \nonumber \\
	& = & 2G (2\eta_{\mu 0} 
	\eta_{\nu 0} - \eta_{\mu \nu} \eta_{00})
	\int d^3y \frac{T^{00}({\bf y})}{|{\bf x}-{\bf y}|}
\label{acca}
\end{eqnarray}

The contracted form of the Einstein equations allows us to compute the
scalar curvature, for any solution, as the trace of the energy-momentum
tensor:
\begin{equation} 
	R(x)=8 \pi G g^{\mu \nu}(x) T_{\mu \nu}(x)
\label{tra} 
\end{equation}
In our case, it is straightforward to check that
$\sqrt{g(x)} g^{00}(x)=1 + o(G^2)$, and so the Einstein-Hilbert action 
computed for the solution of the weak-field equation is
	\begin{equation} 
	S_{zero-mode} = - \frac{1}{2}
	\int d^4x T_{00}(x) + o(G^2)
\label{e17}
\end{equation}
	
Provided the integral of the mass-energy density
vanishes, the action of the field configuration is of
order $G^2$, i.e.\ practically negligible in several cases.
One can give several examples of ``dipolar" sources composed of positive and negative masses, either point-like or in the form of concentric shells, etc.\  \cite{prev}. Some possible applications were mentioned in \cite{prev2}.

\subsection{Strong field}
\label{strong}

After giving an example of zero-modes in perturbation theory,
let us look for dipolar fluctuations with spherical symmetry using the same virtual source approach, but in strong field.
Our starting point is the well-known expression of the Schwarzschild metric 
(from \cite{mtw}, adapted to our conventions)
	\begin{equation}
	d\tau^2 = e^{2\phi(r)} dt^2
	- \frac{dr^2}{1 - 2m(r)G/r} 
	- r^2 \left( d\theta^2 + \sin^2 \theta d\phi^2 \right)
\label{schw}
\end{equation}

The function $m(r)$ represents the mass contained in a 3-sphere of radius $r$:
	\begin{equation}
m(r)=\int_0^r 4\pi {r'}^2 \rho(r') dr'
\end{equation}
where $\rho(r)$ is the mass density in the local inertial frame.

The function $\phi(r)$ appearing in the $g_{00}$ component of the metric is the solution of the system of coupled differential equations
	\begin{equation}
\frac{d\phi(r)}{dr}=\frac{[m(r)+4\pi r^3 p(r)]G}{r[r-2m(r)G]}
\end{equation}
	\begin{equation}
\frac{dp(r)}{dr}=-[\rho(r)+p(r)]\frac{d\phi(r)}{dr}
\end{equation}
where $p$ is the pressure. The second equation is called Oppenheimer-Volkoff equation and amounts to impose the Bianchi identities on the source.
At the external boundary $r_{ext}$ of the mass distribution, $\phi$ satisfies the condition $\phi(r_{ext})= (1/2) \ln(1-2MG/r_{ext})$, where $M$ is the total mass, $M=m(r_{ext})$. 

These relations are often employed in relativistic astrophysical stellar models. In that context, it is also necessary to assign a positive $\rho$; this is not necessary in our case because the source can be un-physical. We try to assign $\rho$ in such a way to obtain metrics for which the integral $\int d^4x \sqrt{g}R$ is zero, by placing a negative-mass shell around the positive core, as done in the weak-field case. 

As a first example, let us consider a core with 
constant positive mass density $\rho_0$, and let $r_s$ be the corresponding Schwarzschild radius 
	\begin{equation}
r_s=\sqrt{\frac{3}{8\pi \rho_0 G}}
\label{rs}
\end{equation}
 						
This is the radius at which a singularity would arise, if the mass distribution was not cut before that point. Let us take an ``inversion radius" $r_{inv}$ and an external radius $r_{ext}$ such that $r_{inv}<r_s<r_{ext}$, and define a density function $\rho(r)$ which is equal to $\rho_0$ up to $r_{inv}$, becomes negative at $r_{inv}$ and then zero from $r_{ext}$ on:
	\begin{equation}
\rho(r)=\rho_0 \varepsilon(r_{inv}-r) \theta(r_{ext}-r)
\label{rho-cost}
\end{equation}
where $\varepsilon$ and $\theta$ are the usual step functions. 
We shall try to define $r_{ext}$ in such a way to satisfy the zero-mode condition $\int d^4x \sqrt{g}R=0$. Note that $r_{inv}$ is arbitrary, provided $r_{inv}<r_s$. If $r_{inv}$ is close to $r_s$, the metric tends to become singular and the volume of the configuration can grow without limit.

In order to write the action density we use the contracted Einstein equations 
(\ref{tra})
and the components of the energy-momentum tensor \cite{mtw}
	\begin{eqnarray}
T^{00}&=&\rho/g_{00} \\
T^{rr}&=&p/g_{rr} \\
T^{\theta \theta}&=&p/r^2 \\
T^{\phi \phi}&=&p/(r^2 \sin^2 \theta)
\end{eqnarray}

We define the adimensional quantities
	\begin{eqnarray}
s&=&r/r_s \\
\tilde{\rho}&=&\rho / \rho_0 \\
\tilde{p}&=&p\rho_0 \\
\tilde{m}&=&2Gm/r_s \\
\tilde{R}&=&\frac{Rr_s^2}{3} 
\label{R-adim}
\end{eqnarray}

We easily obtain in adimensional form
	\begin{equation}
\sqrt{g}=\frac{4\pi s^2 e^\phi}{\sqrt{1-\tilde{m}/s}}; \ \ \ 
\tilde{R}=\tilde{\rho}-3\tilde{p}
\label{lagr}
\end{equation}

The equations for $\phi$ and $p$ are, in adimensional form,
	\begin{equation}
\phi'(s)=\frac{\tilde{m}(s)/s+3s^2\tilde{p}(s)}{2s-\tilde{m}(s)}
\label{phi-adim}
\end{equation}
	\begin{equation}
\tilde{p}'(s)=-[\tilde{p}(s)+\tilde{\rho}(s)]\phi'(s)
\label{p-adim}
\end{equation}

Next we make our treatment more general. Assuming that the virtual density $\rho(r)$ decreases, in absolute value, when $r$ increases, one can obtain several different metrics. The simplest case is that of functions of the form $|\rho(r)| \propto 1/r^k$, with $k\geq 0$. (Lobo \cite{lobo} also employs a source density of this kind, in the context of wormhole solutions, with $k=2$ and without sign inversion. See also \cite{su}.) Eq.s (\ref{rs}), (\ref{rho-cost}), (\ref{R-adim}) and (\ref{phi-adim}) are 
respectively replaced by
	\begin{equation}
r_s=\sqrt{\frac{3-k}{8\pi \rho_0 G}}
\label{rs-gen}
\end{equation}
	\begin{equation}
\rho(r)=\rho_0 \left( \frac{r_0}{r} \right)^k \varepsilon(r_1-r) \theta(r_2-r)
\end{equation}
	\begin{equation}
\tilde{R}=\frac{Rr_s^2}{3-k}
\label{R-adim-gen}
\end{equation}
\begin{equation}
\phi'(s)=\frac{\tilde{m}(s)/s+(3-k)s^2\tilde{p}(s)}{2s-\tilde{m}(s)}
\label{phi-adim2}
\end{equation}

In order to avoid the density divergence in $r=0$, we re-define $\rho(r)$ as a constant from $r=0$ to, say, $r=r_s/2$. It follows that the Schwarzschild singularity does not occur exactly at $r_s$, but slightly on the right, at the point $r=ar_s$, where $a$ is the solution of the algebraic equation
	\begin{equation}
a-a^{3-k}+\frac{k}{3 \cdot 2^{3-k}}=0
\label{eq-per-a}
\end{equation}

With this modification of the density function and this definition of $a$, we finally have in adimensional form
	\begin{equation}
\tilde{\rho}(s)=\theta(a-s) \left[ \theta \left( \frac{1}{2}-s \right) \cdot 2^k
+ \theta \left( s-\frac{1}{2} \right) s^{-k} \right] + \theta(s-a)(-s^{-k})
\end{equation}
	\begin{eqnarray}
\tilde{m}(s)&=&\theta(a-s) \left[ \theta \left( \frac{1}{2}-s \right) 
\left( 1-\frac{k}{3} \right)\cdot 2^k s^3
+ \theta \left( s-\frac{1}{2} \right) \left( s^{3-k}-\frac{k}{3 \cdot 2^{3-k}}
\right) \right] + \\
&  & +\theta(s-a) \left( -s^{3-k}-\frac{k}{3 \cdot 2^{3-k}}
+2a^{3-k} \right) \nonumber
\end{eqnarray}

The coupled equations (\ref{phi-adim}) and (\ref{p-adim}) can now be solved numerically for given values of $k$. It is already apparent from the form of (\ref{p-adim}) that the pressure can in general diverge, because the derivative of $\tilde{p}$ is proportional to $\tilde{p}$ itself, when the density $\tilde{\rho}$ is small. In the usual astrophysical models, $\tilde{p}$ goes rapidly to zero when $r$ increases, and then $\tilde{\rho}$ vanishes, too, at the star surface; in this way, $\tilde{p}$ stays zero outside the star, as it is physically obvious. One would therefore expect that $\tilde{R}=\tilde{\rho}$ outside a certain radius, and that the zero-mode condition can be satisfied by cutting $\tilde{\rho}$. 

\begin{figure}
  \begin{center}
    \includegraphics{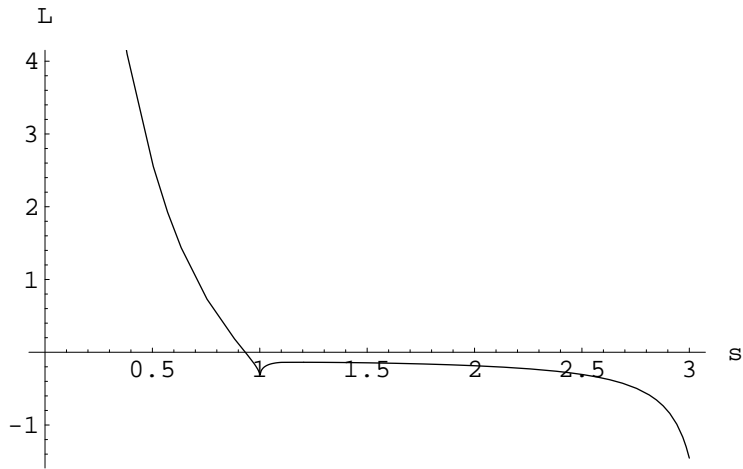}
  \end{center}
  \caption{Solution of the adimensional Oppenheimer-Volkoff equation (\ref{p-adim}) with $k=0.1$, $a = 1.00234$ (eq.\ (\ref{eq-per-a})), density inversion radius $0.999a$, density cutting point $s=1.1$, initial condition $\tilde{p}(0)=10$. The pressure decreases quickly but does not vanish, and diverges after the density cutting point.}
  \label{pressure}
\end{figure}

\begin{figure}
  \begin{center}
    \includegraphics{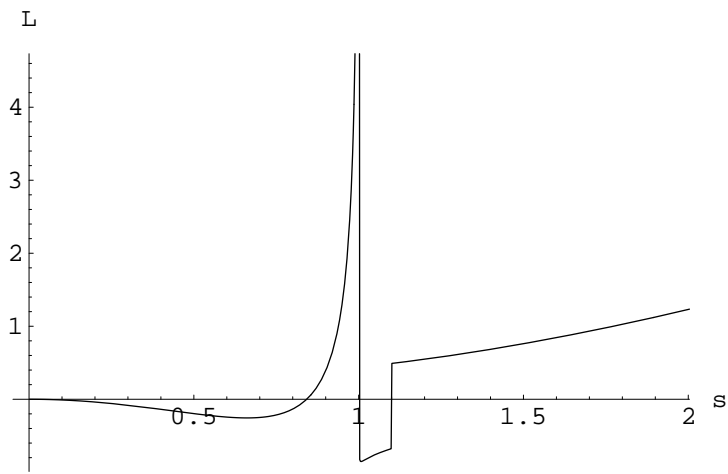}
  \end{center}
  \caption{Adimensional Lagrangian density $\sqrt{g}\tilde{R}$ (eq.\ (\ref{lagr})) with the same parameters as in Fig.\ \ref{pressure}. The scalar curvature diverges due to the pressure component, and the zero-mode condition can clearly not be satisfied.}
  \label{lagrangian}
\end{figure}

In the present case, however, the situation is more tricky. The numerical solutions show that the pressure actually decreases quickly in the positive density core, but keeps non-zero in the outer negative-density shell. If the mass density is cut at $r_{ext}$, the pressure begins to grow exponentially in the negative direction  with a strange self-amplification effect (Fig.\ \ref{pressure}), and the analogy with real sources is completely lost. According to (\ref{lagr}), it follows that the scalar curvature diverges, too, and there is apparently no way to obtain a zero-mode through this method (Fig.\ \ref{lagrangian}). We have tried several values of $k$, in the range $0 <k<2$.

\section{Zero-action modes with constant $g_{00}$}
\label{direct}

Since the virtual-source method appears to be ineffective, we look now for zero-action configurations in a way which is mathematically more straightforward and does not employ the classical concepts of sources and solutions of the Einstein equations. We consider the most general static spherically symmetric metric and try to impose directly the zero-action condition $\int d^4x \sqrt{g} R=0$. Following Weinberg \cite{wei} we write this metric as 
	\begin{equation}
d\tau^2 = B(r)dt^2 - A(r)dr^2 - r^2(d\theta^2+\sin^2 \theta d\phi^2)
\label{sfer-simm}
\end{equation}
where $A$ and $B$ are arbitrary smooth functions.

We add the requirement that outside a certain radius $r_{ext}$, $A$ and $B$ take the Schwarzschild form, namely
	\begin{equation}
B(r) = \left( 1 - \frac{2GM}{r} \right); \ \ \ 
A(r) = \left( 1 - \frac{2GM}{r} \right)^{-1} \ \ \ {\rm for} \ r \geq r_{ext}
\label{ABesterni}
\end{equation}

This requirement serves two purposes:

(1) It allows to give a physical meaning to these configurations, seen from the outside, as mass-energy fluctuations of strength $M$. For $r>r_{ext}$ their scalar curvature is zero.

(2) The York-Hawking-Gibbons boundary term of the action is known to take in this case the form $I=-M\int dt$ \cite{eqg}. It is therefore constant for our perturbed solution with the same $M$ and $L \neq 0$ (see below).

Weinberg gives an useful expression for the Ricci tensor as a function of $A$ and $B$: 	
	\begin{eqnarray}
R_{rr} & = & \frac{B''}{2B} - \frac{1}{4}b(a+b) - \frac{a}{r} \\
R_{\theta \theta} & = & -1 + \frac{r}{2A}(-a+b) + \frac{1}{A} \\
R_{\phi \phi} & = & \sin^2 \theta R_{\theta \theta} \\
R_{tt} & = & -\frac{B''}{2A} + \frac{B'}{4A}(a+b) - \frac{B'}{rA}
\label{comp-ricci}
\end{eqnarray}
where $a=A'/A$, $b=B'/B$ and all functions depend only on $r$.

In order to obtain the curvature scalar $R$, we contract $R_{\mu \nu}$ with the components of $g^{\mu \nu}$. (Note that Weinberg's convention is $d\tau^2 = - g_{\mu \nu} dx^\mu dx^\nu$, therefore $g_{rr}=A$, $g_{tt}=-B$, $g_{\theta \theta}=r^2$, 
$g_{\phi \phi}=r^2 \sin^2 \theta$.) We obtain
	\begin{equation}
R = -\frac{R_{tt}}{B} + \frac{R_{rr}}{A} + 2 \frac{R_{\theta \theta}}{r^2}
\label{Rscalare}
\end{equation}

Given any two smooth trial functions $A$ and $B$, it is straightforward to compute $\sqrt{g} R$. The problem is, that we do not have any control on the outcome, because the expression for $R$ is complicated and highly non-linear; so imposing the integral condition $\int d^4x \sqrt{g} R=0$ appears to be exceedingly difficult.

We do find a set of solutions, however, if we make the drastic simplification $g_{00}=B=const$. The scalar curvature multiplied by the volume element becomes in this case
	\begin{equation}
L = \sqrt{g}R = -8\pi \sqrt{|BA|} \left( \frac{rA'}{A^2} + 1 -\frac{1}{A} \right)
\label{Lsemplif}
\end{equation}
Apart from the Newton constant $G$, $L$ is the lagrangian density of the Einstein-Hilbert action, computed for this particular metric.

Let us fix arbitrarily a reference radius $r_{ext}$, and introduce reduced coordinates
$s=r/r_{ext}$. Define an auxiliary function $\alpha=A^{-1}$. Regarding $L(s)$ as known, eq.\ (\ref{Lsemplif}) becomes an explicit first-order differential equation for $\alpha$ 
	\begin{equation}
\alpha' = \frac{1}{s} - \frac{\alpha}{s} + \frac{L\sqrt{|\alpha|}}{8\pi s 
\sqrt{|B|}}
\label{eq-diff-alpha}
\end{equation}

The boundary conditions (\ref{ABesterni}) are written
	\begin{equation}
B(s \geq 1) = \left( 1 - \frac{\tilde{M}}{s} \right); \ \ \ 
A(s \geq 1) = \left( 1 - \frac{\tilde{M}}{s} \right)^{-1}
\label{condizioneAB}
\end{equation}
where $\tilde{M}$ is a free parameter, the total mass in reduced units: 
$\tilde{M}=2GM/r_{ext}$. In the following we shall take $\tilde{M}<0$, in order to avoid singularities. For $r<r_{ext}$, we have $B=B(1)=1-\tilde{M}$.

It is interesting to note that putting $L=0$ in eq.\ (\ref{eq-diff-alpha}) we can easily find an exact solution, ie a non-trivial static metric with $R=0$. Namely,
if $1-\alpha>0$, then $\alpha=1-e^{const}/s$, which does not satisfy the boundary condition; if $1-\alpha<0$, then $\alpha=1+e^{const}/s$, implying 
$e^{cost}=-\tilde{M}$. The resulting $g_{rr}$ component has the same form on the left and on the right of $s=1$, namely
	\begin{equation}
g_{rr} = \left( 1 + \frac{|\tilde{M}|}{s} \right)^{-1}
\label{grr-statica}
\end{equation}
while $g_{00}$ is constant and equal to $(1 + |\tilde{M}|)$ for $s<1$, and is equal to $(1 + |\tilde{M}|/s)$ for $s\geq 1$.

\begin{figure}
  \begin{center}
    \includegraphics{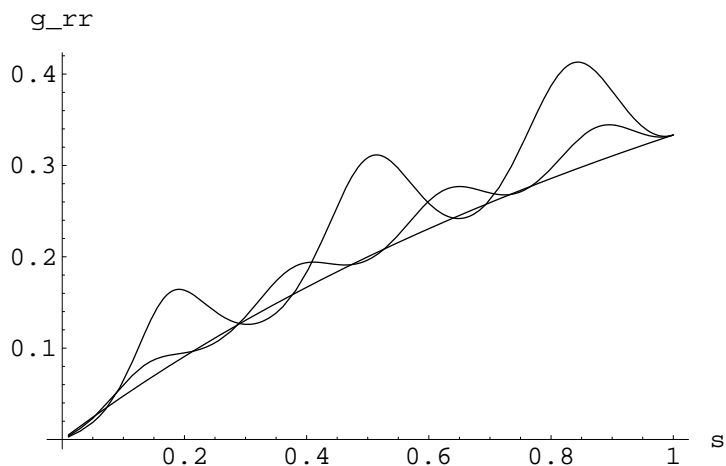}
  \end{center}
  \caption{The $g_{rr}$ metric components of: a zero-mode with $\tilde{M}=-2$ and null scalar curvature (lower line); a zero-mode with the same $\tilde{M}$ and adimensional lagrangian $L=8\sin(6\pi s)$ (upper line, with three peaks); a zero-mode with the same $\tilde{M}$ and $L=4\sin(8\pi s)$ (intermediate line, with four peaks).}
  \label{grr}
\end{figure}

For large $|\tilde{M}|$ and small $L$, the last term in eq.\ (\ref{eq-diff-alpha}) is a small perturbation. Since $\alpha$ never diverges and $\alpha^{-1}$ does not appear in the equation, the perturbed solution is not very different from (\ref{grr-statica}) (see below). For values of $|\tilde{M}|$ of order 1 or smaller, the equation can be integrated numerically. If we choose a function $L(s)$ with null integral on the interval $(0,1)$, we obtain a metric which is a zero-mode of the action but not of the lagrangian, like those we unsuccessfully looked for through the virtual source method. Take, for instance, $L(s)=L_0 \sin(2\pi n s)$, with $n$ integer (Fig.\ \ref{grr}).

In conclusion, we have found a family of metrics with null scalar curvature, depending on a continuous parameter $\tilde{M}$, which in our finite-dimensional analogy represent the level line $R=0$. Furthermore, we have built a set of metrics close to the latter, by solving eq.\ (\ref{eq-diff-alpha}) with $L$ arbitrary but having null integral in the interval $(0,1)$. These metrics do not have zero scalar curvature, but still have null action. They make up a full-dimensional subset of the functional space; in order to show this, let us re-write eq. (\ref{Lsemplif}) in operator form as follows (remember that $\alpha(s)=g_{rr}^{-1}(s)$ and $B=g_{00}=const.$):
	\begin{equation}
L(s)= -\frac{8\pi \sqrt{B}}{\sqrt{|\alpha|}} (s \alpha'(s)-\alpha(s)+1) \equiv
T \alpha(s)
\label{def-T}
\end{equation}
Eq.\ (\ref{eq-diff-alpha}) can be written in operator form as
	\begin{equation}
\alpha(s) = T^{-1} L(s)
\label{def-T-1}
\end{equation}

Consider functions $\alpha(s)$ and $L(s)$ in the interval $(0,1)$. Let $E_0$ be the set of the metrics $\alpha$ satisfying the boundary condition (\ref{ABesterni}). Let $E_1$ be the image of $E_0$ through $T$, namely $T:E_0 \longrightarrow E_1$. Any $L(s) \epsilon E_1$ can be expressed in Fourier series as follows:
	\begin{equation}
L(s) = \sum_{n=0}^\infty \left[ a_n \sin(2\pi ns) + b_n \cos(2\pi ns) \right]
\label{fourier}
\end{equation}
Eliminate from $E_1$ the functions with $a_0 \neq 0$. The remaining ensemble,
$E_2$, contains only functions with null integral in $(0,1)$. We make the conjecture that

(1) The ensemble $E_2$ is infinite-dimensional.

(2) The back-image $T^{-1}(E_2)$ (contained in $E_0$) is infinite-dimensional.

A rigorous proof of this conjecture is not trivial. The conjecture could only be false, however, if the operator $T^{-1}$ was quite singular, projecting $E_2$ into a finite-dimensional space. But this in not the case, as can be seen by solving the equation explicitily in the perturbative approximation of small $L$. If $\tilde{M} \gg 1$, after replacing $\alpha$ and $B$ in the square root with their unperturbed solution (\ref{grr-statica}), eq.\ (\ref{eq-diff-alpha}) simplifies to
	\begin{equation}
\alpha' = \frac{1}{s} - \frac{\alpha}{s} + \frac{L}{8\pi s^{3/2}}
\label{eq-diff-alpha-pert}
\end{equation}
which can be solved to give
	\begin{equation}
\alpha(s) = \frac{1}{s} \left[ \int \left( 1+\frac{L(s')}{8\pi \sqrt{s'}}
\right) ds' + const. \right]
\label{soluz-pert}
\end{equation}
We see that in this case the operator $T^{-1}$ is linear and its nucleus contains only the function $L(s)=-8\pi \sqrt{s}$. Our conjecture is therefore true. On the other hand, if $\tilde{M} \ll 1$, 
eq.\ (\ref{eq-diff-alpha}) takes the form
	\begin{equation}
\alpha' = \frac{1}{s} - \frac{\alpha}{s} + \frac{L}{8\pi \sqrt{s}}
\label{eq-diff-alpha-pert2}
\end{equation}
whose solution is
	\begin{equation}
\alpha(s) = \frac{1}{s} \left[ \int \left( 1+\frac{\sqrt{s'}L(s')}{8\pi}
\right) ds' + const. \right]
\label{soluz-pert2}
\end{equation}
Also in this case the operator $T^{-1}$ is linear, its nucleus containing only the function $L(s)=-\frac{8\pi}{\sqrt{s}}$, and our conjecture is  true.

\section{Conclusions}

Although a generally accepted theory of quantum gravity does not yet exist, the aim of this work was to give some insight into its vacuum state. To this end, we considered the formal path integral over all metric configurations, weighed by the factor $e^{iS/\hbar}$. This factor oscillates in general very fast, and the most probable configurations are those which make the phase $S/\hbar$ constant or stationary. The metrics which make the action stationary are, of course, the classical configurations. A WKB-like expansion near these configurations in powers of $\hbar$ has been previously considered by Hawking and others in the context of the Euclidean functional integral approach. Usually, in quantum mechanics or quantum field theory the configurations with stationary action are local minima; so if the system leaves them, the phase of the weight factor starts to oscillate more and more quickly, until the one-loop approximation fails to be valid and a pure quantum regime is entered. In quantum gravity, on the contrary, the non-positivity of the action implies that it is possible to depart from the stationary point along configurations at which the action takes the same value it takes at the stationary point. These configurations actually have vanishing Lagrangian density. Each of them can be furthermore perturbed, or ``deformed", in such a way as to obtain metrics which do not have zero Lagrangian at each point, but have zero total action due to the cancellation between positive and negative contributions from different regions of space.

These zero-modes are very different from the classical solutions of Einstein equations and can not be obtained employing virtual sources or starting from wormhole-like metrics. It was actually to be expected that a full quantum feature of spacetime could not be recovered by ``quantizing the classical solutions". The static and spherically symmetric zero-modes described in this work are probably just members of a much wider class. They are characterized by a large negative mass. They are not singular and outside a certain radius they look like Schwarzschild solutions. These metrics are physically realized as vacuum fluctuations and are therefore homogeneously distributed in space and time, but could possibly manifest their presence if some cause of inhomogeneity is introduced, as it happens in electromagnetism for the Casimir effect. Finally, they could give a relevant contribution to the cosmological term. These latter issues will be addressed in future work.

\bigskip

{\bf Acknowledgement.} This work was partially supported by the Institut fuer Gravitationsforschung, Waldaschaff, Germany.

\end{document}